\documentclass[aps,twocolumn,prabib,amsmath,amssymb,superscriptaddress,notitlepage]{revtex4-1}
\usepackage{graphics}
\usepackage{bm}
\usepackage{dcolumn}
\usepackage{natbib}
\usepackage{epstopdf}
\usepackage{float}
\usepackage{graphicx}
\usepackage{epsfig}
\usepackage[pdfstartview=FitH]{hyperref}
\usepackage{color}
\usepackage{appendix}
\usepackage{ulem}

\newcommand{\rr}{{\bf r}}

\begin{document}
\title{Bosonic Halperin fractional quantum Hall effect at filling factor $\nu=2/5$}
\author{Tian-Sheng Zeng}
\thanks{These authors contributed equally to this work.}
\affiliation{Department of Physics, School of Physical Science and Technology, Xiamen University, Xiamen 361005, China}
\author{Liangdong Hu}
\thanks{These authors contributed equally to this work.}
\affiliation{Zhejiang University, Hangzhou 310027, China and \\
 School of Science, Westlake University, 18 Shilongshan Road, Hangzhou 310024, China}
\author{W. Zhu}
\affiliation{Key Laboratory for Quantum Materials of Zhejiang Province, School of Science, Westlake University, 18 Shilongshan Road, Hangzhou 310024, China}
\date{\today}
\begin{abstract}
Quantum Hall effects with multicomponent internal degrees of freedom facilitate the playground of novel emergent topological orders.
Here, we explore the correlated topological phases of two-component hardcore bosons at a total filling factor $\nu=2/5$ in both lattice Chern band models and Landau level continuum model
under the interplay of intracomponent and intercomponent repulsions.
We give the numerically theoretical demonstration of the emergence of two competing distinct fractional quantum Hall states: Halperin (441) fractional quantum Hall effect and Halperin (223) fractional quantum Hall effect.
We elucidate their topological features including the degeneracy of the ground state and fractionally quantized topological Chern number matrix. Finally, we discuss scenarios related to phase transition between them when intercomponent nearest-neighbor coupling is tuned from weak to strong in topological checkerboard lattice.
\end{abstract}
\maketitle

\section{Introduction}

Multicomponent systems provide an avenue for realizing a tremendous amount of physics that have
no analogue in one-component systems, which also results in a question of fundamental interest: what kind of topological order can emerge in a microscopic interacting model.
Two-component fractional quantum Hall (FQH) effect is such an example,
which is conjectured to produce new incompressible ground states beyond Laughlin states.
The earlier theoretical studies of an electron gas in spin-unpolarized and bilayer (or double quantum wells) systems such as AlGaAs,
exemplified by integer quantum Hall state at $\nu=1$~\cite{Wen1992,Moon1995,Yang1996} and fractional quantum Hall state at $\nu=1/2$~\cite{Yoshioka1988,Yoshioka1989,He1991,He1993} (note that whereas for single layer, the other competing candidates could be either a composite fermion liquid in the lowest Landau level or a Moore-Read state in the second Landau level),
can be described as Halperin's two-component $(mmn)$ wave functions with the $\mathbf{K}=\begin{pmatrix}
m & n\\
n & m\\
\end{pmatrix}$ matrix~\cite{Halperin1983}.
Recently, experimental observations in graphene have greatly expanded our understanding of
a widespread zoology of multicomponent FQH effects~\cite{Bolotin2009,Dean2011,Liu2019,Li2019} with a SU(4) theoretical generalization of Halperin's wave function to these FQH effects in graphene~\cite{Goerbig2007},
and at the same time inspired the current study of multicomponent FQH effect in bosonic systems as the counterparts.

Compared to the fermionic systems, the study of two-component bosonic quantum Hall effects affords us much ongoing knowledge and yet demands further attention. Besides FQH states of spinless bosons at sequential filling factors in early days~\cite{Cooper2020}, it was realized that there exists an intimate theoretical generalizations of the Halperin's wave functions to clustered spin-singlet quantum Hall states~\cite{Ardonne1999,Reijnders2002,Ardonne2005,Grosfeld2009,Ardonne2011,Davenport2012,Hormozi2012} for bosons in the lowest Landau level with an internal degree of freedom like pseudospin,
and even a classification scheme for bosonic symmetry-protected topological (SPT) phases with no intrinsic topological order for multicomponent bosons~\cite{Chen2013},
such as two-component bosonic integer quantum Hall liquid with the associated $\mathbf{K}=\begin{pmatrix}
0 & 1\\
1 & 0\\
\end{pmatrix}$ matrix in the presence of mutual flux attachment~\cite{Moller2009,Senthil2013,Lu2012}, followed by the numerical hunt for two-component bosonic fractional quantum Hall states at various filling factors where several proposals of new quantum Hall structures mentioned above, have been examined i.e. spin-singlet FQH states at $\nu=2k/3$ and integer quantum Hall state at $\nu=2$~\cite{Grass2012,Furukawa2012,Hu2013,Furukawa2013,Regnault2013,Grass2014,Furukawa2017,Nakagawa2017}
based on either strong synthetic gauge fields or topological bands with high Chern number~\cite{Sterdyniak2015,Moller2015} in cold atomic neutral systems, and the possible emergence of tunneling-induced Moore-Read quantum Hall state at $\nu=1$ from two copies of Laughlin $\nu=1/2$ FQH state~\cite{Zhu2015,Liu2016} that are not available in electronic FQH effect.
Nevertheless, the vague outline of multicomponent bosonic systems discloses a large uncharted theoretical area of two-dimensional bosonic topological order~\cite{Wen2016} to be explored,
aside from the experimental interest.

With these motivations, in this work, we are concerned with the internal correlation structures of two-component bosonic FQH effects at fixed filling factor $\nu=2/5$
where a convincing theoretical evidence is still lacking.
We will consider both the continuum Landau level and topological flat band models. Early pioneering theoretical studies~\cite{Lukin2005,Lukin2007,Palmer2006,Palmer2008,Kapit2010} suggest that Laughlin-like FQH state of neutral atoms can be simulated in lattice models in the presence of artificial gauge field which plays the role of magnetic field acting on charged electrons.
While Landau level problem directly relates to the problems in the high magnetic field, 
topological flat band models dubbed ``Chern insulators'' are becoming an excitingly new platform~\cite{Haldane1988}
for studying the quantum Hall effect at zero magnetic field,
along with lots of experimental interests in topological Hofstadter-Harper~\cite{Aidelsburger2013,Miyake2013,Mancini2015,Stuhl2015}
and Haldane-honeycomb~\cite{Jotzu2014} bands for cold atom realization,
topological moir\'e minibands for twisted multilayer graphene~\cite{Wolf2019,Chittari2019,Chen2020}
and other flat bands with nontrivial topology in geometrically frustrated lattice~\cite{Kang2020}.
The emergence of FQH effect in topological flat bands (namely ``fractional Chern insulators''~\cite{Sun2011,Sheng2011,Neupert2011,Wang2011,Tang2011,Regnault2011}) requires a demanding understanding of the internal topological structure of interacting fractionalised phases,
where an integer valued symmetric $\mathbf{K}$ matrix was proposed to characterize different topological orders for Abelian multicomponent systems according to the Chern-Simons theory
~\cite{Wen1992a,Wen1992b,Blok1990a,Blok1990b,Blok1991}. Indeed, at partial fillings $\nu=1/(kC+1)$ (odd $k$
for hardcore bosons and even $k$ for spinless fermions) in topological flat bands with higher Chern numbers $C$,
there fractionalised Abelian $C$-color-entangled states host a close relationship to $C$-component FQH states
~\cite{LBFL2012,Wang2012r,Yang2012,Sterdyniak2013,Wang2013,Moller2015,YLWu2013,YLWu2014,YHWu2015,Behrmann2016},
where the general one-to-one correspondence is built up based on the classification of their $\mathbf{K}$ matrices from the inverse of Chern number matrix for these gapped topological phases
~\cite{Zeng2017,Zeng2018,Zeng2019,Zeng2020},
where the quantized intercomponent drag Hall transport is identified as a primary evidence for the emergence of exotic correlated many-body topological states in multicomponent systems~\cite{Liu2019,Li2019}. Together with synthetic magnetic gauge fields in cold atomic neutral systems, these related progresses, thus enable new relevant prospects for the study of two-component bosonic FQH effects in both lattice and continuum models, which is the focus of our work.

The main findings of the present work is that, we characterize two robust bosonic FQH state at filling factor $\nu=2/5$: (a) in the spin-$S_z$-rotation symmetric limit $V_{\uparrow\downarrow}/V_{\uparrow\uparrow}\ll1$, the system is Halperin (441) FQH state with $\mathbf{K}=\begin{pmatrix}
4 & 1\\
1 & 4\\
\end{pmatrix}$, and (b) in the spin-SU(2)-symmetric limit $V_{\uparrow\downarrow}/V_{\uparrow\uparrow}\simeq1$, the system is Halperin (223) FQH state with $\mathbf{K}=\begin{pmatrix}
2 & 3\\
3 & 2\\
\end{pmatrix}$, as presented below ($V_{\sigma\sigma'}$ denotes
the nearest-neighbor interaction between bosons of $\sigma$-component and those of $\sigma'$-component ).
Through the state-of-the-art methods, including ground state degeneracy,
topological Chern number matrix on the torus manifold and fractional charge pumping, chiral edge excitations on the cylindrical geometry,
we determine their topological nature.

The remainder of the paper is organized as follows.
In Sec.~\ref{model}, we give a description of the model Hamiltonian of interacting two-component bosons in both lattice and continuum models, with focuses on two typical topological lattice models,
such as $\pi$-flux checkerboard and Haldane-honeycomb lattices.
In Sec.~\ref{H441} and Sec.~\ref{H223}, we study the many-body ground states of these two-component bosonic systems at $\nu=2/5$ in both lattice and continuum models,
and present detailed numerical demonstration of the $\mathbf{K}$ matrix classification of competing quantum Hall states based on the ground state degeneracy,
Chern number matrix. In Sec.~\ref{edge}, we discuss chiral edge physics from the view of bulk entanglement spectrum. In Sec.~\ref{phasetransition}, we qualitatively discuss the phase transition between competing quantum Hall states base on our ED calculation.
Finally, we conclude in Sec.~\ref{summary} with the prospect of investigating competing phases in two-component bosonic systems.

\section{Model and Method}\label{model}

\subsection{Topological lattice models}

Here, we utilize both exact diagonalization (ED) and density-matrix renormalization group (DMRG) to study the low-energy properties of the Hamiltonian for two-component hardcore bosons with pseudospin degrees of freedom via intercomponent and intracomponent interactions at a total filling $\nu=2/5$ in topological flat bands, and elucidate the physical mechanism of the competing Halperin FQH effects. The Hamiltonian built on topological $\pi$-flux checkerboard (CB) and Haldane-honeycomb (HC) lattices, is given by
\begin{align}
  H_{CB}=&\!\sum_{\sigma}\!\Big[-t\!\!\sum_{\langle\rr,\rr'\rangle}\!e^{i\phi_{\rr'\rr}}b_{\rr',\sigma}^{\dag}b_{\rr,\sigma}
  -\!\!\!\!\sum_{\langle\langle\rr,\rr'\rangle\rangle}\!\!\!t_{\rr,\rr'}'b_{\rr',\sigma}^{\dag}b_{\rr,\sigma}\nonumber\\
  &-t''\!\!\!\sum_{\langle\langle\langle\rr,\rr'\rangle\rangle\rangle}\!\!\!\! b_{\rr',\sigma}^{\dag}b_{\rr,\sigma}+H.c.\Big]+V_{int},\label{cbl}\\
  H_{HC}=&\!\sum_{\sigma}\!\Big[-t\!\!\sum_{\langle\rr,\rr'\rangle}\!\! b_{\rr',\sigma}^{\dag}b_{\rr,\sigma}-t'\!\!\sum_{\langle\langle\rr,\rr'\rangle\rangle}\!\!e^{i\phi_{\rr'\rr}}b_{\rr',\sigma}^{\dag}b_{\rr,\sigma}\nonumber\\
  &-t''\!\!\sum_{\langle\langle\langle\rr,\rr'\rangle\rangle\rangle}\!\!\!\! b_{\rr',\sigma}^{\dag}b_{\rr,\sigma}+H.c.\Big]+V_{int},\label{hcl}
\end{align}
where $b_{\rr,\sigma}^{\dag}$ is the hardcore bosonic creation operator of pseudospin $\sigma=\uparrow,\downarrow$ at site $\rr$, $\langle\ldots\rangle$,$\langle\langle\ldots\rangle\rangle$ and $\langle\langle\langle\ldots\rangle\rangle\rangle$ denote the nearest-neighbor, the next-nearest-neighbor, and the next-next-nearest-neighbor pairs of sites, respectively. We take the on-site and nearest-neighbor interactions,
\begin{align}
  V_{int}=&U\sum_{\rr}n_{\rr,\uparrow}n_{\rr,\downarrow}
  +V_{\uparrow\uparrow}\sum_{\sigma=\uparrow,\downarrow}\sum_{\langle\rr,\rr'\rangle}n_{\rr',\sigma}n_{\rr,\sigma} \nonumber\\
 &+V_{\uparrow\downarrow}\sum_{\langle\rr,\rr'\rangle}\big(n_{\rr',\uparrow}n_{\rr,\downarrow}+n_{\rr',\downarrow}n_{\rr,\uparrow}\big)
\end{align}
where $n_{\rr,\sigma}$ is the particle number operator of pseudospin $\sigma$ at site $\rr$. In what follows, we impose the tunnel couplings $t'=0.3t,t''=-0.2t,\phi=\pi/4$ for checkerboard lattice, while $t'=0.6t,t''=-0.58t,\phi=2\pi/5$ for honeycomb lattice~\cite{Wang2011,Wang2012}, and choose the interaction strengths $U=\infty,V_{\uparrow\uparrow}/t=100$ with variable $V_{\uparrow\downarrow}$ ranging from $V_{\uparrow\downarrow}/V_{\uparrow\uparrow}=0$ to SU(2)-symmetric $V_{\uparrow\downarrow}/V_{\uparrow\uparrow}=1$.

For finite system sizes with translational symmetry, $\nu=\nu_{\uparrow}+\nu_{\downarrow},\nu_{\uparrow}=2N_{\uparrow}/N_s=\nu_{\downarrow}=2N_{\downarrow}/N_s=1/5$, where $N_s=2\times N_x\times N_y$ is the total number of sites, and $N_{\sigma}$ is the particle number of pseudospin. The energy states are labeled by the total momentum $K=(K_x,K_y)$ in units of $(2\pi/N_x,2\pi/N_y)$ in the Brillouin zone. For cylindrical system sizes, we exploit both infinite DMRG and finite DMRG on the cylindrical geometry with open boundary condition in the $x$ direction and periodic boundary condition in the $y$ direction, and the bond dimension of DMRG is kept up to $M=6500$ here.

\subsection{Landau continuum model}

\begin{figure}[b]
  \centering
  \includegraphics[height=1.6in,width=3.4in]{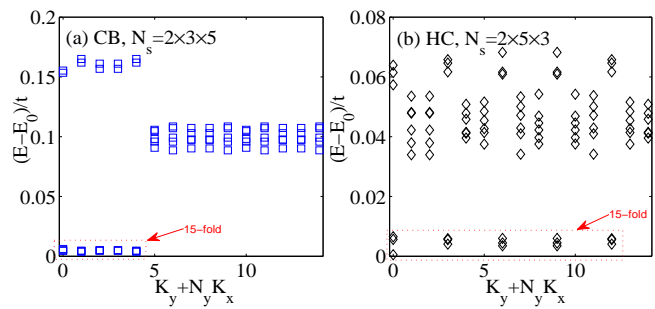}
  \caption{\label{energy} (Color online)
  Numerical ED results for the low energy spectrum of two-component bosonic systems $\nu=2/5$ with $U(1)\times U(1)$-symmetry for spin-$S_z$-rotation symmetric interaction $V_{\uparrow\downarrow}/V_{\uparrow\uparrow}=0$ in different topological lattice models (a) checkerboard lattice and (b) honeycomb lattice. The red dotted box indicates the ground state degeneracy.}
\end{figure}

Meanwhile, as an unabridged physical counterpart, we also utilize ED to investigate the continuous model on periodic torus spanned by vectors $\vec L_1 = L \hat e_x$ and $\vec L_2 = L \hat e_y$. The single particle Hamiltonian, which can be realized in rapidly rotating bosonic gases~\cite{Cooper2008}, is written as:
\begin{equation}\label{Hamiltonian_LL}
H_0(\mathbf{A}) = \frac12 \sum_{a=x,y}D_a(\mathbf{A}) D_a(\mathbf{A}).  \\
\end{equation}
Where $D_a(\mathbf{A})= p_a-qA_a = -i\hbar \nabla_a +qA_a$ is the canonical momentum in magnetic field and $\mathbf{A}=(-By,0)$ is the vector potential.
The total number of magnetic fluxes $N_{s}$ penetrating the torus is given by the Landau level degeneracy
$N_{s}=\frac{\vec{L}_1 \times \vec{L}_2}{2\pi \ell^2}=\frac{ L^2}{2\pi \ell^2}$,
where the magnetic length $\ell=\sqrt{\hbar/qB}$ is taken as the unit of the length.
The eigenstates of Eq.~\ref{Hamiltonian_LL} are called Landau levels. Here, we study the FQH state in the lowest Landau level
with $N_s$-fold degenerate orbits~\cite{Yoshioka1983}:
\begin{equation}\label{wavefunction_LL}
\psi_j = \frac{\mathrm{e}^{-\frac{y^2}{2\ell^2}}}{\sqrt{\pi^{\frac 12}L\ell}}\sum_{n\in Z}e^{-\pi N_{s} (n+\frac{j}{N_{s}})^2 +i2\pi N_{s}(n+\frac{j}{N_{s}})\frac{z}{L}}
\end{equation}
where $j=0,1,\cdots,N_{s}-1$. In the lowest Landau level, the two-body interactions between particles can be written as:
\begin{align}\label{H_twobody}
U_{\sigma_1 \sigma_2} &=\!\sum_{j_1,j_2,j_3,j_4} A_{j_1,j_2,j_3,j_4} b_{\sigma_1 j_1}^\dagger b_{\sigma_2 j_2}^\dagger b_{\sigma_2 j_3} b_{\sigma_1 j_4}, \\
A_{j_1,j_2,j_3,j_4}&=\!\!\! \sum_{n_x,n_y}\! \frac{U(|\mathbf{q}|)}{2L^2}e^{-\frac12 \mathbf{q}^2\ell^2- i2\pi \frac{n_y}{N_\phi}(j_1-j_3)}\delta_{j_1,j_2,j_3,j_4}    \nonumber
\end{align}
where the symbol $\delta_{j_1,j_2,j_3,j_4}=\delta^{\mathrm{mod} N_{s}}_{j_1-j_4,n_x}\delta^{\mathrm{mod} N_{s}}_{j_1+j_2,j_3+j_4}$, $U(|\mathbf{q}|) = \int\int \mathrm{d}x\mathrm{d}y U(|\mathbf{r_1}-\mathbf{r_2}|)\mathrm{e}^{-i\mathbf{q}\cdot (\mathbf{r_1}-\mathbf{r_2})}$ is the Fourier
transformation of particle-particle interaction $U(|\mathbf{r_1}-\mathbf{r_2}|)$ and $\mathbf{q}=\frac{2\pi}{L}(n_x,n_y)$ is the reciprocal lattice vector.
The Dirac symbol $\delta^{\mathrm{mod} N_{s}}_{i,j}$ means that the equivalence between $i$ and $j$ is defined by modulo $N_{s}$ and $\sigma$ represents the layer index
(or pseudospin degree of freedom).

We consider the double-layer system with all bosons in each layer limited to the lowest Landau level, and the interaction Hamiltonian between bosons contains three parts:
\begin{align}\label{H_twolayer}
H = U_{\uparrow \uparrow}  + U_{\uparrow \downarrow}  +  U_{\downarrow \uparrow}  +  U_{\downarrow \downarrow},
\end{align}
where $U_{\sigma_1 \sigma_2}$ is defined by Eq.~\ref{H_twobody}, and $U_{\uparrow \uparrow}, U_{\downarrow \downarrow}$ is the intra-layer interaction in the first and second layer while $U_{\uparrow \downarrow}$ denotes the inter-layer interactions between different layers.

\section{Halperin (441) Fractional Quantum Hall Effect}\label{H441}

In this section, we first delve into the topological ground state degeneracies for the interaction
$V_{\uparrow\downarrow}/V_{\uparrow\uparrow}=0$,
where the SU(2)-symmetric repulsion is broken down to spin-$S_z$-rotation symmetry, and discuss its stability.
Next we give a detailed demonstration of the $\mathbf{K}$ matrix classification from the inverse of the Chern number matrix $\mathbf{C}$ based on ED calculation of topologically invariant Chern number and DMRG simulation of drag charge pumping in the periodic parameter plane.

\begin{figure}[b]
  \centering
  \includegraphics[height=1.25in,width=3.4in]{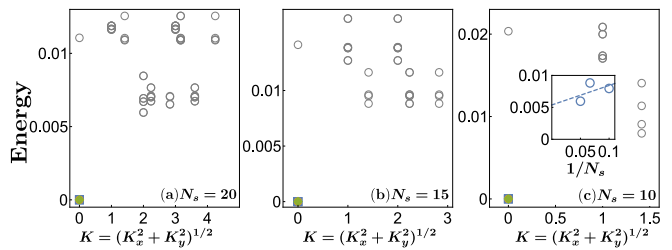}
  \caption{\label{441_hu_energy} (Color online) Numerical ED results for the low energy spectrum of two-component bosonic systems $\nu=2/5$ with $U(1)\times U(1)$-symmetry in Landau continuum model. There are three-fold degenerate ground states in a single irreducible Brillouin zone. The inset depicts the finite size scaling of energy gap which persists in the thermodynamic limit.}
\end{figure}

\subsection{Ground state degeneracy}

The key property of the existence of topological fractionalised ordered phases lies in their ground state degeneracies.
For Halperin $(mmn)$ quantum Hall state, the ground state degeneracy is given by the determinant of the $\mathbf{K}$ matrix.
Thus, first we demonstrate the ground state degeneracy on periodic lattice in different interacting regimes.
As shown in Figs.~\ref{energy}(a) and~\ref{energy}(b) for different topological systems in the limit $V_{\uparrow\downarrow}/V_{\uparrow\uparrow}=0$,
there exists a low-energy manifold with fifteen-fold degenerate ground states, which are separated from higher energy levels by a robust gap.

\begin{figure}[t]
  \includegraphics[height=2.6in,width=3.4in]{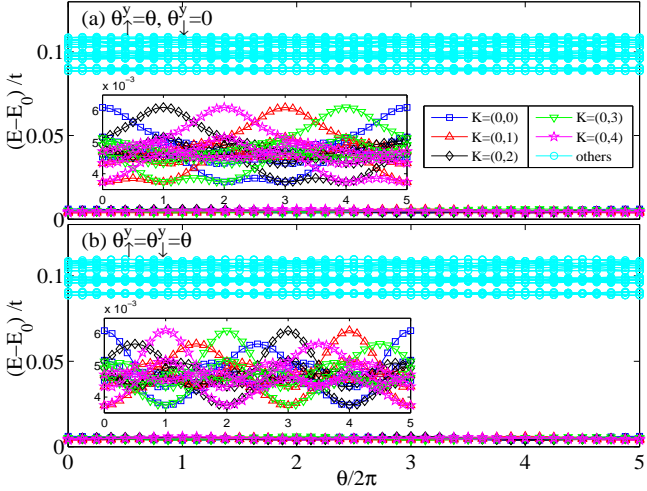}
  \caption{\label{flux} (Color online) Numerical ED results for the low energy spectral flow of two-component bosonic systems $\nu=2/5,N_s=2\times3\times5$ with $U(1)\times U(1)$-symmetry for the interaction $V_{\uparrow\downarrow}/V_{\uparrow\uparrow}=0$ in topological checkerboard lattice under the insertion of two types of flux quanta (a) $\theta_{\uparrow}^{y}=\theta,\theta_{\downarrow}^{y}=0$ and (b) $\theta_{\uparrow}^{y}=\theta_{\downarrow}^{y}=\theta$.}
\end{figure}

Similarly, we also study the ground state degeneracy in the continuum Landau level model as an intuitive physical understanding.
The energy states are labeled by the pseudo-momentum $K=(K_x,K_y)$ in units of $(2\pi/L, 2\pi/L)$ in the irreducible Brillouin zone~\cite{Haldane1985}. Since the translational
symmetry of center of mass, there are five equivalent irreducible Brillouin zones. Thus, for FQH states in $\nu=2/5$, we have at least five-fold degenerated ground states.
When we choose Haldane pseudopotentials $v_0=v_2=1.0$ for $U_{\uparrow \uparrow}(U_{\downarrow \downarrow})$ and $v_0=1.0$ for $U_{\uparrow \downarrow}$, the ground states with exact zero
energy host three-fold degeneracy in each irreducible Brillouin zone (namely 15-fold for five equivalent irreducible Brillouin zones when the center-of-mass translation invariance is taken into account) as shown in Fig.~\ref{441_hu_energy}, consistent with the theoretical
prediction of Halperin (441) state.
In the continuum model, we can access three different system sizes. For all of these system sizes,
the ground state degeneracy is robust, and the energy gap separating the ground state manifold and the excitations
is finite which is nonzero in the finite-size scaling (see the inset of Fig.~\ref{441_hu_energy}(a)).

Further, in order to demonstrate the topological equivalence of these ground states,
we calculate the low energy spectra flux under the insertion of flux quanta,
by utilizing twisted boundary conditions $\psi(\rr_{\sigma}+N_{\alpha})=\psi(\rr_{\sigma})\exp(i\theta_{\sigma}^{\alpha})$ where $\theta_{\sigma}^{\alpha}$
is the twisted angle for pseudospin $\sigma$ particles in the $\alpha$ direction.
For $V_{\uparrow\downarrow}/V_{\uparrow\uparrow}=0$, as shown in Fig.~\ref{flux}(a), these fifteen-fold ground states evolve into each other without mixing with the higher levels,
and the system returns back to itself upon the insertion of five flux quanta for both $\theta_{\uparrow}^{\alpha}=\theta_{\downarrow}^{\alpha}=\theta$
and $\theta_{\uparrow}^{\alpha}=\theta,\theta_{\downarrow}^{\alpha}=0$.
The robustness of degenerate ground states reveals the existence of universal internal structures with the behavior of fractional quantization.

In essence, on both lattice model and continuum models, we have identified the robustness of
fifteen-fold ground state degeneracy on the torus geometry at $\nu=2/5$. This Halperin state survives in both lattice Chern band models and Landau level continuum model, which demonstrate its robustness for future detecting.

\subsection{Chern number matrix}

\begin{figure}[t]
  \centering
  \includegraphics[height=1.55in,width=3.4in]{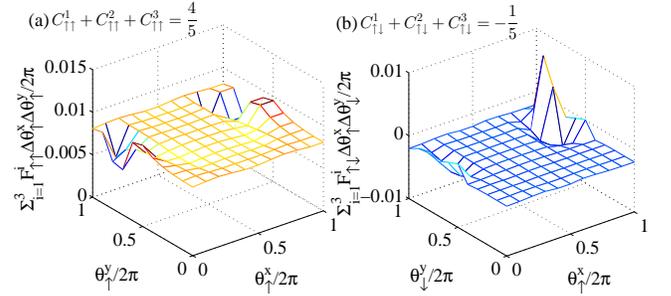}
  \caption{\label{berry} (Color online) Numerical ED results for Berry curvatures $F_{\sigma\sigma'}^{xy}\Delta\theta_{\sigma}^{x}\Delta\theta_{\sigma'}^{y}/2\pi$ of the three ground states at $K=(0,0)$ of two-component bosonic systems $N_{\uparrow}=N_{\downarrow}=5,N_s=2\times3\times5$ for the interaction $V_{\uparrow\downarrow}/V_{\uparrow\uparrow}=0$ in topological checkerboard lattice in different parameter planes. (a) $(\theta_{\uparrow}^{x},\theta_{\uparrow}^{y})$ and (b) $(\theta_{\uparrow}^{x},\theta_{\downarrow}^{y})$.}
\end{figure}

Following the above discussion, we continue to analyze the fractionally topological quantization, characterized by the Chern number matrix $\mathbf{C}=\begin{pmatrix}
C_{\uparrow\uparrow} & C_{\uparrow\downarrow} \\
C_{\downarrow\uparrow} & C_{\downarrow\downarrow} \\
\end{pmatrix}$ of the many-body ground state wavefunction $\psi$ for interacting systems~\cite{Sheng2003,Sheng2006}.
In the parameter plane $(\theta_{\sigma}^{x},\theta_{\sigma'}^{y})$,
the matrix elements are defined by $C_{\sigma\sigma'}=\int d\theta_{\sigma}^{x}d\theta_{\sigma'}^{y}F_{\sigma\sigma'}(\theta_{\sigma}^{x},\theta_{\sigma'}^{y})/2\pi$ with the Berry curvature
\begin{align}
  F_{\sigma\sigma'}(\theta_{\sigma}^{x},\theta_{\sigma'}^{y})=\mathbf{Im}\left(\langle{\frac{\partial\psi}{\partial\theta_{\sigma}^x}}|{\frac{\partial\psi}{\partial\theta_{\sigma'}^y}}\rangle
-\langle{\frac{\partial\psi}{\partial\theta_{\sigma'}^y}}|{\frac{\partial\psi}{\partial\theta_{\sigma}^x}}\rangle\right).\nonumber
\end{align}
By numerically calculating the Berry curvatures using $m\times m$ mesh Wilson loop plaquette in the boundary phase space with $m\geq10$,
one can obtain the quantized topological invariant $C_{\sigma\sigma'}$ of the gapped ground states at momentum $K$, and $C_{\uparrow\uparrow}=C_{\downarrow\downarrow},C_{\uparrow\downarrow}=C_{\downarrow\uparrow}$ under the prescribed symmetry $b_{\rr,\uparrow}\leftrightarrow b_{\rr,\downarrow}$.
In the ED study of finite system size, we find that for $V_{\uparrow\downarrow}/V_{\uparrow\uparrow}=0$,
the three ground states at momentum $K=(0,0)$ couple with each other under the flux thread, and host the average many-body Chern numbers $\frac{1}{3}\sum_{i=1}^3C^i_{\uparrow\uparrow}=0.2621\simeq4/15$ and $\frac{1}{3}\sum_{i=1}^3C^i_{\uparrow\downarrow}=-0.0641\simeq-1/15$ for $10\times10$ mesh points with the absolute error around $0.002$, as indicated in Figs.~\ref{berry}(a) and~\ref{berry}(b) respectively.

\begin{figure}[t]
  \centering
  \includegraphics[height=1.6in,width=3.4in]{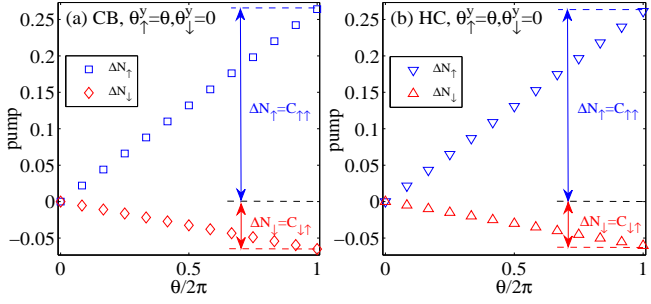}
  \caption{\label{pump} (Color online) Quantized charge transfers for two-component bosonic systems $\nu=2/5$ with $U(1)\times U(1)$-symmetry on the infinite $N_y=5$ cylinder for spin-$S_z$-rotation symmetric interaction $V_{\uparrow\downarrow}/V_{\uparrow\uparrow}=0$ under the insertion of flux quantum $\theta_{\uparrow}^{y}=\theta,\theta_{\downarrow}^{y}=0$ in one component for (a) checkerboard lattice and (b) honeycomb lattice.}
\end{figure}

As topological Chern number is related to the Hall transport response~\cite{Niu1985},
we further calculate the charge pumping induced by the Berry curvature once the flux quantum is adiabatically inserted on infinite cylinder systems using DMRG~\cite{Gong2014}.
Numerically we cut the infinite cylinder into left-half and right-half parts along the $x$ direction,
and obtain the net charge transfer of the pseudospin $\sigma$ from the right side to the left side by calculating the the evolution of
$N_{\sigma}(\theta_{\sigma'}^{y})=tr[\widehat{\rho}_L(\theta_{\sigma'}^{y})\widehat{N}_{\sigma}]$
as a function of $\theta_{\sigma'}^{y}$ (here $\widehat{\rho}_L$ the reduced density matrix of the left part, classified by the quantum numbers $\Delta Q_{\uparrow},\Delta Q_{\downarrow}$).
For $V_{\uparrow\downarrow}/V_{\uparrow\uparrow}=0$, as shown in Fig.~\ref{pump}(a) and~\ref{pump}(b),
we get fractionally quantized charge transfer in different topological lattice models with pumping values
$\Delta N_{\uparrow}=N_{\uparrow}(2\pi)-N_{\uparrow}(0)\simeq C_{\uparrow\uparrow}=4/15,\Delta N_{\downarrow}=N_{\downarrow}(2\pi)-N_{\downarrow}(0)\simeq C_{\downarrow\uparrow}=-1/15$
upon threading one flux quantum $\theta_{\uparrow}^y$ of the pseudospin $\uparrow$ with $\theta_{\downarrow}^y=0$ for two-component bosons with the absolute error around $0.001$.

In view of quantized topological invariants, our ED and DMRG studies establish the essential diagnosis of distinct topological orders,
benefitting from the merit of the well-defined Chern number matrix of the gapped ground state, namely
\begin{align}
  \mathbf{C}=\frac{1}{15}\begin{pmatrix}
4 & -1\\
-1 & 4\\
\end{pmatrix}, \mathbf{K}=\mathbf{C}^{-1}
\end{align}
for $V_{\uparrow\downarrow}/V_{\uparrow\uparrow}\ll1$ where the system falls into Halperin's $(441)$ FQH state.

\section{Halperin (223) Fractional Quantum Hall Effect}\label{H223}

In this section, we turn to analyze the topological ground state properties in topological checkerboard lattice for the interaction
$V_{\uparrow\downarrow}/V_{\uparrow\uparrow}=1$, where the SU(2)-symmetric repulsion is restored. Similarly, we would present numerical proofs of the $\mathbf{K}$ matrix classification from the inverse of the Chern number matrix $\mathbf{C}$ based on ED calculation of topologically invariant Chern number and DMRG simulation of drag charge pumping in the periodic parameter plane.

\begin{figure}[t]
  \centering
  \includegraphics[height=1.7in,width=3.4in]{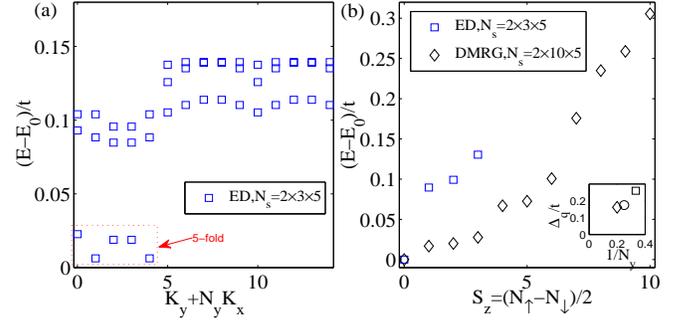}
  \caption{\label{energy223} (Color online)
  Numerical results for two-component bosonic systems $\nu=2/5$ with $U(1)\times U(1)$-symmetry for spin-SU(2)-symmetric interaction $V_{\uparrow\downarrow}/V_{\uparrow\uparrow}=1$ in topological checkerboard lattice model. (a) ED study of the low energy spectrum with $N_s=2\times3\times5$ and (b) ED and DMRG studies of polarization in different spin sectors. The red dotted box indicates the ground state degeneracy and the inset depicts the scaling of charge-hole gap $\Delta_q$ for different cylinder systems $N_y=3,4,5,N_s=2\times10\times N_y$.}
\end{figure}

\begin{figure}[b]
  \centering
  \includegraphics[height=1.75in,width=3.4in]{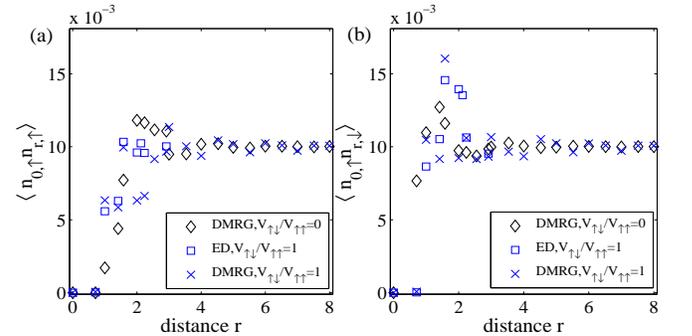}
  \caption{\label{pair} (Color online)
  Numerical results for two-component bosonic systems $\nu=2/5$ with $U(1)\times U(1)$-symmetry in topological checkerboard lattice model. (a) Pair correlation inside the same component for different interactions $V_{\uparrow\downarrow}/V_{\uparrow\uparrow}$ and (b) Pair correlation between different components for different interactions $V_{\uparrow\downarrow}/V_{\uparrow\uparrow}$. For ED, the system size $N_s=2\times3\times5$, and for DMRG, the infinite cylinder system with $N_y=5$ is used. The distance $r=\sqrt{x^2+y^2}$ is measured in units of lattice constant, and the original part of $\langle n_{\uparrow,0}n_{\uparrow,r}\rangle$ at $r=0$ is subtracted.}
\end{figure}

\subsection{Ground state degeneracy}

In contrast to the limit $V_{\uparrow\downarrow}/V_{\uparrow\uparrow}\ll1$, we find that in the SU(2)-symmetric limit $V_{\uparrow\downarrow}/V_{\uparrow\uparrow}=1$,
there exists a low-energy manifold with five-fold degenerate ground states, which are separated from higher energy levels by a robust gap, as shown in Figs.~\ref{energy223}(a). In the continuum model, however we find no numerical evidence of a well-defined topological ground manifold by trying different combinations of SU(2)-symmetric pseudopotentials, which is shown to be possibly unstable within the plasma picture~\cite{Gail2008}. Therefore we would consider its possible unstability against other competing phases in our lattice models, like spin-polarized ferromagnetic order or phase separation. Figure~\ref{energy223}(b) shows the charge-hole gap for different cylinder system sizes in fixed spin sector $S_z=(N_{\uparrow}-N_{\downarrow})/2=0$ and the ground state energy in different spin sectors $S_z=(N_{\uparrow}-N_{\downarrow})/2$. Our finite DMRG simulation of the charge-hole gap $\Delta_q=(E_0(N_{\uparrow}+1,N_{\downarrow})+E_0(N_{\uparrow}-1,N_{\downarrow})-2E_0(N_{\uparrow},N_{\downarrow}))/2$ gives a positive value, which serves as a characteristics of an incompressible FQH liquid. Further, the lowest ground state energy always falls into the spin sector $S_z=0$ with $E_0(S_z=0)<E_0(S_z\neq0)$. This observation meets the the fact that Halperin (223) FQH state should be an SU(2) spin-singlet phase, whose $\mathbf{K}$ matrix is related to the Cartan matrix of the Lie algebra SU(2) by a special linear group transformation~\cite{Wen1992a}.

\begin{figure}[t]
  \centering
  \includegraphics[height=2.6in,width=3.4in]{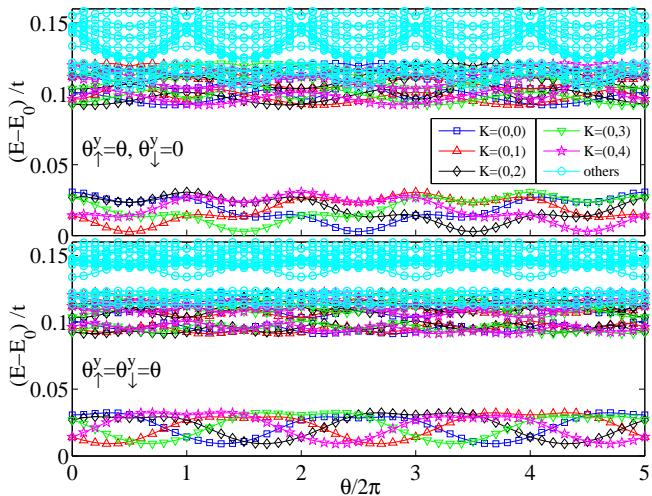}
  \caption{\label{flux223} (Color online) Numerical ED results for the low energy spectral flow of two-component bosonic systems $\nu=2/5,N_s=2\times3\times5$ with $U(1)\times U(1)$-symmetry for SU(2)-symmetric interaction $V_{\uparrow\downarrow}/V_{\uparrow\uparrow}=1$ in topological checkerboard lattice under the insertion of two types of flux quanta $\theta_{\uparrow}^{y}=\theta,\theta_{\downarrow}^{y}=0$ (upper panel) and $\theta_{\uparrow}^{y}=\theta_{\downarrow}^{y}=\theta$ (lower panel).}
\end{figure}

Moreover, to exclude a tendency towards phase separation, we plot the pair-correlation profile of the ground state wavefunction $|\psi\rangle$ with $S_z=0$, as shown in Figs.~\ref{pair}(a) and~\ref{pair}(b). The uniform density correlation at long distances not only demonstrates a homogeneous mixture of bosons of both components, but also eliminates the potential translational symmetry breaking phase. When changed rom $V_{\uparrow\downarrow}/V_{\uparrow\uparrow}=0$ to $V_{\uparrow\downarrow}/V_{\uparrow\uparrow}=1$, the short-range pair correlations $\langle n_{\uparrow,0}n_{\uparrow,r}\rangle$ enhances while $\langle n_{\uparrow,0}n_{\downarrow,r}\rangle$ is suppressed near the distance $r=1$ due to strong nearest-neighbor interaction $V_{\uparrow\downarrow}$. Also we confirm that the local density has a balanced occupation $\langle n_{\rr,\uparrow}\rangle=\langle n_{\rr,\downarrow}\rangle$. Compared to the continuum model, our discrete lattice model provides an alternative way for engineering certain unusual FQH effects.

Similar to Sec.~\ref{H441}, in order to demonstrate their topological equivalence, we calculate the low energy spectra flux under the insertion of flux quanta,
by utilizing twisted boundary conditions $\psi(\rr_{\sigma}+N_{\alpha})=\psi(\rr_{\sigma})\exp(i\theta_{\sigma}^{\alpha})$ where $\theta_{\sigma}^{\alpha}$
is the twisted angle for pseudospin $\sigma$ particles in the $\alpha$ direction.
For $V_{\uparrow\downarrow}/V_{\uparrow\uparrow}=1$, as shown in Fig.~\ref{flux223}, these five-fold ground states evolve into each other without mixing with the higher levels,
and the system returns back to itself upon the insertion of five flux quanta for both $\theta_{\uparrow}^{\alpha}=\theta_{\downarrow}^{\alpha}=\theta$
and $\theta_{\uparrow}^{\alpha}=\theta,\theta_{\downarrow}^{\alpha}=0$.
The robustness of inherent five-fold degenerate ground states in the SU(2)-symmetric limit $V_{\uparrow\downarrow}/V_{\uparrow\uparrow}=1$,  implies the existence of novel internal topological structures, in distinction from the behavior of Halperin (441) FQH states in the limit $V_{\uparrow\downarrow}/V_{\uparrow\uparrow}=0$.

\begin{figure}[t]
  \centering
  \includegraphics[height=1.45in,width=3.4in]{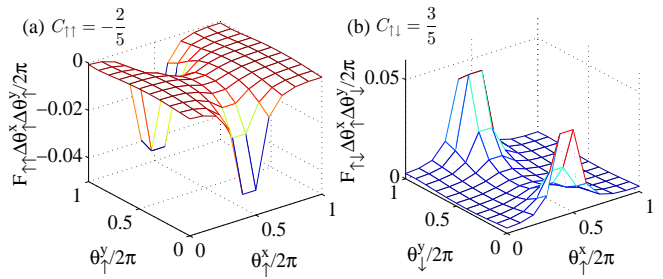}
  \caption{\label{berry223} (Color online) Numerical ED results for Berry curvatures $F^{xy}\Delta\theta_{\sigma}^{x}\Delta\theta_{\sigma'}^{y}/2\pi$ of the single ground state at $K=(0,0)$ of two-component bosonic systems $N_{\uparrow}=N_{\downarrow}=5,N_s=2\times3\times5$ for SU(2)-symmetric interaction $V_{\uparrow\downarrow}/V_{\uparrow\uparrow}=1$ in topological checkerboard lattice in different parameter planes. (a) $(\theta_{\uparrow}^{x},\theta_{\uparrow}^{y})$ and (b) $(\theta_{\uparrow}^{x},\theta_{\downarrow}^{y})$.}
\end{figure}

\subsection{Chern number matrix}

Similar to Sec.~\ref{H441}, to further clarify the distinctive topological phase in the SU(2)-symmetric limit $V_{\uparrow\downarrow}/V_{\uparrow\uparrow}=1$, we calculate the Chern number matrix $\mathbf{C}=\begin{pmatrix}
C_{\uparrow\uparrow} & C_{\uparrow\downarrow} \\
C_{\downarrow\uparrow} & C_{\downarrow\downarrow} \\
\end{pmatrix}$ of the many-body ground state wavefunction $\psi$ for interacting systems in the parameter plane $(\theta_{\sigma}^{x},\theta_{\sigma'}^{y})$.
In the ED study of finite system size, for single ground state at momentum $K=(0,0)$, we plot the corresponding Berry curvatures $F_{\uparrow\uparrow},F_{\uparrow\downarrow}$ which vary smoothly, as indicated in Figs.~\ref{berry223}(a) and~\ref{berry223}(b) respectively. We obtain the topological invariants $C_{\uparrow\uparrow}\simeq-0.3923,C_{\uparrow\downarrow}\simeq0.6077$ for $10\times10$ mesh points, which are remarkably close to fractionally quantized ideal values $C_{\uparrow\uparrow}=-2/5,C_{\uparrow\downarrow}=3/5$ with the relative error around one percentage.

\begin{figure}[t]
  \centering
  \includegraphics[height=1.6in,width=3.3in]{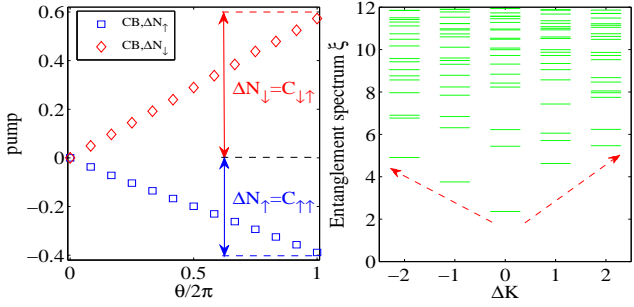}
  \caption{\label{es2} (Color online) Left panel: Quantized charge transfers for two-component bosonic systems $\nu=2/5$ with $U(1)\times U(1)$-symmetry on the infinite $N_y=5$ cylinder for SU(2)-symmetric interaction $V_{\uparrow\downarrow}/V_{\uparrow\uparrow}=1$ under the insertion of flux quantum $\theta_{\uparrow}^{y}=\theta,\theta_{\downarrow}^{y}=0$. Right panel: Non-chiral edge nature for two-component bosonic systems $\nu=2/5$ with $U(1)\times U(1)$-symmetry for SU(2)-symmetric interaction $V_{\uparrow\downarrow}/V_{\uparrow\uparrow}=1$ on the infinite $N_y=5$ cylinder in the typical charge sector $\Delta Q_{\uparrow}=0,\Delta Q_{\downarrow}=-1$ for topological checkerboard lattice.}
\end{figure}

For larger system sizes, we continue to calculate the charge pumping on infinite cylinder systems utilizing the same DMRG simulation of adiabatic flux insertion in Sec.~\ref{H441}.
For $V_{\uparrow\downarrow}/V_{\uparrow\uparrow}=1$, as shown in the left panel of Fig.~\ref{es2},
we get fractionally quantized charge transfer in topological checkerboard lattice model, encoded by
$\Delta N_{\uparrow}=N_{\uparrow}(2\pi)-N_{\uparrow}(0)=-0.3888\simeq C_{\uparrow\uparrow}=-2/5,\Delta N_{\downarrow}=N_{\downarrow}(2\pi)-N_{\downarrow}(0)=0.5712\simeq C_{\downarrow\uparrow}=3/5$
upon threading one flux quantum $\theta_{\uparrow}^y$ of the pseudospin $\uparrow$ with $\theta_{\downarrow}^y=0$ for two-component bosons.

Therefore, combined with the results of ground state degeneracy, we can establish the $\mathbf{K}$ matrix classification of Halperin $(223)$ FQH state, derived from the inverse of the Chern number matrix of the gapped ground states, namely
\begin{align}
  \mathbf{C}=\frac{1}{5}\begin{pmatrix}
-2 & 3\\
3 & -2\\
\end{pmatrix}, \mathbf{K}=\mathbf{C}^{-1}
\end{align}
for $V_{\uparrow\downarrow}/V_{\uparrow\uparrow}\simeq1$.

\section{Entanglement Spectroscopy of Edge Physics}\label{edge}

Moreover, we analyze the edge physics of these topological ordered phases according to the bulk-edge correspondence,
especially the chiral Luttinger liquid character which can be used to characterize the topological orders in the different FQH states~\cite{Wen1992edge,Wen1995}.
The edge chirality is determined by the signs of the eigenvalues $m\pm n$ of the $\mathbf{K}=\begin{pmatrix}
m & n\\
n & m\\
\end{pmatrix}$ matrix.

\begin{figure}[t]
  \centering
  \includegraphics[height=1.65in,width=3.4in]{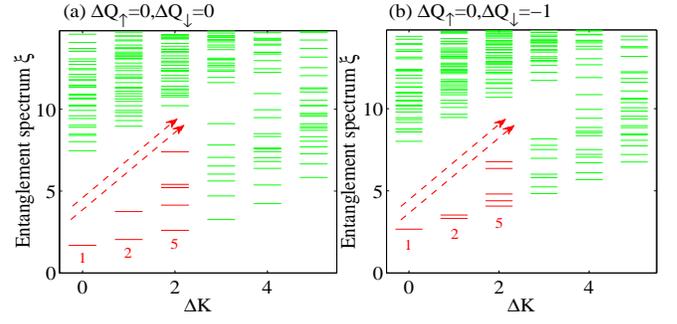}
  \caption{\label{es} (Color online) Chiral edge mode identified from the momentum-resolved entanglement spectrum for two-component bosonic systems $\nu=2/5$ with $U(1)\times U(1)$-symmetry for spin-$S_z$-rotation symmetric interaction $V_{\uparrow\downarrow}/V_{\uparrow\uparrow}=0$ on the infinite $N_y=6$ cylinder in the typical charge sectors (a) $\Delta Q_{\uparrow}=\Delta Q_{\downarrow}=0$ and (b) $\Delta Q_{\uparrow}=0,\Delta Q_{\downarrow}=-1$. The horizontal axis shows the relative momentum $\Delta K=K_y-K_{y}^{0}$ (in units of $2\pi/N_y$). The numbers below the red dashed line label the level counting: $1,2,5,\cdots$.}
\end{figure}

\begin{figure}[b]
  \centering
  \includegraphics[height=1.9in,width=3.4in]{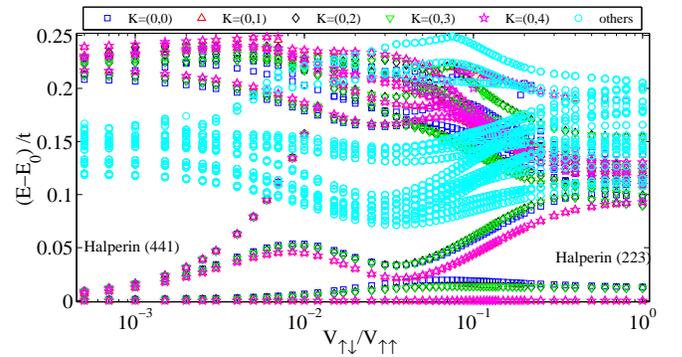}
  \caption{\label{phase} (Color online) Numerical ED results for the low energy evolution of two-component bosonic systems $\nu=2/5,N_s=2\times3\times5$ with $U(1)\times U(1)$-symmetry in topological checkerboard lattice as a function of $V_{\uparrow\downarrow}/V_{\uparrow\uparrow}$.}
\end{figure}

Here we harness the low-lying momentum-resolved entanglement spectrum to identify the topological nature on the infinite cylinder based on the general relationship between the entanglement spectrum and the spectrum of a physical edge state~\cite{Li2008,Qi2012}:
(i) for $m>n$, two propagating chiral modes in the same direction are obtained~\cite{Zeng2017},
(ii) for $m<n$, two chiral modes propagate in the opposite directions, i.e.
bosonic integer quantum Hall phase~\cite{He2015}, (iii) for $m=n$, only one chiral mode is left, i.e.
Halperin (111) exciton phase~\cite{Zeng2020exciton}. For $V_{\uparrow\downarrow}/V_{\uparrow\uparrow}=1$, as shown in the right panel of Fig.~\ref{es2}, we observe two oppositely moving branches of low-lying excitations, consistent with the non-chiral nature of Halperin (223) FQH state. However, for $V_{\uparrow\downarrow}/V_{\uparrow\uparrow}=0$,
Figures~\ref{es}(a) and~\ref{es}(b) depict two parallel forward-moving branches of low-lying excitations,
matching with the level counting $1,2,5,\cdots$ of WZW conformal field description for two gapless free bosonic edge theories of Halperin (441) FQH state.

\section{Phase Transition}\label{phasetransition}

Finally, since two different competing topological orders emerge in the same clean system, one fundamental question is about the possible phase transition nature between them with varying $V_{\uparrow\downarrow}/V_{\uparrow\uparrow}$. When tuning $V_{\uparrow\downarrow}/V_{\uparrow\uparrow}=0$ continuously to $V_{\uparrow\downarrow}/V_{\uparrow\uparrow}=1$, we plot the evolution of the low energy spectrum in Fig.~\ref{phase}. Even a moderate intercomponent interaction $V_{\uparrow\downarrow}/V_{\uparrow\uparrow}\sim0.01$ tends to destroy the topological degenerate manifold of Halperin (441) FQH states by diminishing the protecting energy gap. For finite system sizes, our ED study gives a continuous crossover among these degenerate ground states, with a possible level repulsion at the intermediate regime $V_{\uparrow\downarrow}/V_{\uparrow\uparrow}\sim0.04$. However, a much detailed study (taking account of finite size effects) is particularly demanded to make clear conclusive statements about the nature of the transition. It is still possible that in the thermodynamic limit the system can then undergo a simple level crossing, resulting in a first-order transition due to the mismatched nature of topological order~\cite{McDonald1996} or there is an intermediate crossover regime in between, which is beyond our current computability.

\section{Conclusion and Outlook}\label{summary}

So far we have introduced both microscopic lattice model and Landau continuum model that realize
two-component bosonic FQH effects at the filling $\nu=2/5$, including Halperin (441) quantum Hall states in the extremely spin-$S_z$-rotation symmetric interacting limit and Halperin (223) quantum Hall states in the nearly SU(2)-symmetric interacting limit.
Using ED and DMRG simulations, we find that the ground state shows several characteristic topological properties in connection to the $\mathbf{K}$ matrix:
(i) the ground state degeneracy, (ii) fractional topological Chern number in relation to Hall conductance and (iii) chiral edge modes. A tiny interaction perturbation trying to restore the SU(2) symmetry would quickly destroy the stability of this Halperin (441) FQH effect. Our study thus offers a unique perspective of competing quantum Hall physics in the clean system with multiple components, and might furthermore serve as a promising paradigm for engineering this mechanism just by tuning intercomponent and intracomponent interactions using artificial topological bands
in future experiments on cold atomic gases~\cite{Cooper2019} or rapidly rotating bosonic gases.

\section*{Acknowledgements}

T.S.Z. particularly thanks D. N. Sheng for inspiring discussions and  prior collaborations on multicomponent quantum Hall effects. W.Z. thanks Zhao Liu for helpful discussions. This work is supported by the funding of Westlake University, and the NSFC under Grant No. 11974288, the Key R$\&$D Program of Zhejiang Province China (2021C01002) (L.H., W.Z.). This work is also partially supported by the NSFC under Grant No. 12074320 (T.S.Z.).

\end{document}